\documentclass[twocolumn,superscriptaddress,floatfix,preprintnumbers, nofootinbib,hyperref]{revtex4-2} 
\pdfoutput=1
\usepackage[colorlinks=true,breaklinks=true]{hyperref}
\usepackage[normalem]{ulem}
\usepackage[utf8]{inputenc}
\hypersetup{allcolors=[rgb]{0.0 0.0 0.6},linkcolor=[rgb]{0.75 0.05 0.05}}
\usepackage{amsmath,amssymb}
\usepackage{epsfig}  
\usepackage{graphicx}   
\usepackage{slashed}       
\usepackage{tikz}
\usepackage{tikz-feynman}
\usepackage{url}
\usepackage{color}
\usepackage{multirow}
\usepackage{comment}
\usepackage{amssymb}

\DeclareMathOperator{\GeV}{GeV}
\DeclareMathOperator{\eV}{eV}

\DeclareMathOperator{\MeV}{MeV}

\DeclareMathOperator{\s}{s}

\newcommand{\beq}{\begin{equation}}
\newcommand{\eeq}{\end{equation}}

\allowdisplaybreaks

\bibpunct{[}{]}{,}{n}{}{,}

\pgfarrowsdeclare{X}{X}
{
  \pgfutil@tempdima=0.3pt%
  \advance\pgfutil@tempdima by.25\pgflinewidth%
  \pgfutil@tempdimb=5.5\pgfutil@tempdima\advance\pgfutil@tempdimb by.5\pgflinewidth%
  \pgfarrowsleftextend{+-\pgfutil@tempdimb}
  \pgfarrowsrightextend{0pt}
}
{
  \pgfutil@tempdima=0.3pt%
  \advance\pgfutil@tempdima by.25\pgflinewidth%
  \pgfsetdash{}{+0pt}
  \pgfsetroundcap
  \pgfsetmiterjoin
  \pgfpathmoveto{\pgfqpoint{-5.5\pgfutil@tempdima}{-6\pgfutil@tempdima}}
  \pgfpathlineto{\pgfqpoint{5.5\pgfutil@tempdima}{6\pgfutil@tempdima}}
  \pgfpathmoveto{\pgfqpoint{-5.5\pgfutil@tempdima}{6\pgfutil@tempdima}}
  \pgfpathlineto{\pgfqpoint{5.5\pgfutil@tempdima}{-6\pgfutil@tempdima}}
  \pgfusepathqstroke 
}
\makeatother

\begin{document}
\title{Composite heavy axion-like dark matter}

\author{Pierluca Carenza}
\affiliation{The Oskar Klein Centre, Department of Physics, Stockholm University, Stockholm 106 91, Sweden}

\author{Roman Pasechnik}
\affiliation{
 Department of Physics, Lund University
 SE-223 62 Lund, Sweden
}

\author{Zhi-Wei Wang}
\affiliation{School of Physics, The University of Electronic Science and Technology of China,\\
 No.~2006, Xiyuan Ave, West Hi-Tech Zone, Chengdu, China}

\smallskip

\begin{abstract}
We propose a novel class of Dark Matter (DM) candidates in the form of a heavy composite Axion-Like Particle (ALP) with highly suppressed electromagnetic interactions populating vast yet unexplored domains in the ALP parameter space. This is achieved for the first time in the simplest dark confining gauge theory yielding a new composite glueball ALP (GALP) DM coupling-mass relation found in terms of two distinct fundamental scales -- the large dark fermion mass scale and the dynamical scale of dark confinement. The presence of a heavy fermion portal between the visible (photons) and dark (GALPs) sectors ensures a strong radiative suppression of the GALP-photon coupling naturally without any fine-tuning. The observable features of heavy GALP DM in a minimal realization are controlled by only three physical parameters. Our work paves the road for a novel research field exploring the theory and phenomenology of composite ALPs in multi-messenger astrophysics and cosmology.
\end{abstract}

\maketitle

\emph{\textbf{Introduction.}---} Over past decades, several new fundamental particles have been proposed as possible particle Dark Matter (DM) candidates, each with unique properties and signatures. The most popular DM candidates are typically separated into two different classes of models based upon their mass scales -- wave-like (light DM) and particle-like (heavy DM). In the first class of models in the light (sub-eV) mass range, the QCD axion and Axion-Like Particles (ALPs), collectively referred to as `axions', are considered among the promising DM candidates~\cite{Preskill:1982cy,Abbott:1982af,Dine:1982ah,OHare:2024nmr,Adams:2022pbo}. Axions may appear in various extensions of the Standard Model (SM), such as QCD axions that address the strong CP problem~\cite{Peccei:1977hh,Peccei:1977ur,Weinberg:1977ma,Wilczek:1977pj,DiLuzio:2020wdo}, or ALPs emerging in model inspired by string theory such as the ``string axiverse''~\cite{Witten:1984dg,Demirtas:2018akl,Demirtas:2021gsq,Conlon:2006tq,Svrcek:2006yi,Cicoli:2012sz,Arvanitaki:2009fg}. Numerous ongoing and planned experimental searches aim to detect axion-type DM particles, primarily through their photon coupling~\cite{Sikivie:1983ip,Graham:2015ouw,Sikivie:2020zpn}. While several models with the axion mass above keV have been proposed~\cite{Svrcek:2006yi,Cicoli:2012sz,Halverson:2019kna} (for a review, see e.g.~Ref.~\cite{DiLuzio:2020wdo}), heavy axions are never considered to be viable DM candidates due to their efficient decay into lighter SM particles. In the second class of models well above the eV scale, usually at the GeV scale or beyond, Weakly Interacting Massive Particles (WIMPs) and heavy WIMP-like states remain attractive DM candidates~\cite{Steigman:1984ac,Bertone:2004pz,Leane:2018kjk,Cirelli:2024ssz}, which gained considerable attention for their role in explaining the observed $\gamma$-ray excess from the Milky Way center~\cite{Goodenough:2009gk,Izaguirre:2014vva,Fermi-LAT:2017opo,Leane:2019xiy}. The WIMP annihilation signatures are searched for not only in energetic $\gamma$-fluxes but also in cosmic rays and neutrinos~\cite{Cirelli:2024ssz} placing a bound on the mass \mbox{$m_{\rm WIMP}\gtrsim(20-100)$~GeV}, or even lower~\cite{Nollett:2013pwa,Nollett:2014lwa}, depending on the annihilation channels~\cite{MAGIC:2016xys,Leane:2018kjk}.

There has been a great effort to probe parameter spaces for both ALP and WIMP types of DM, placing severe exclusion bounds on both classes of models. In view of the negative results of the searches, it becomes highly relevant to probe novel yet unexplored classes of DM models, with a much larger parameter space sharing some of the basic characteristics of both ALPs and WIMPs. In this work, we propose a new type of DM candidates that serve this purpose featuring both the ALP type of interactions with the visible sector and WIMP-like mass scales, and beyond, thus, significantly expanding the parameter space available for ALP DM searches.

\emph{\textbf{Challenges of ALPs.}---}Consider the case of an axion field $a$ coupled only to photons via the effective 5-dimensional operator
\begin{equation}
\mathcal{L} =- \frac{1}{4}g_{a\gamma} a\, F_{\mu\nu} \tilde F^{\mu\nu}  \,,
\label{eq:ALP}
\end{equation}
where $g_{a\gamma}$ is the axion-photon coupling, $F_{\mu\nu}$ is the electromagnetic field strength tensor and $\tilde F^{\mu\nu}$ is its dual. In this case, the axion $a\to\gamma\gamma$ decay rate,
\begin{equation}\label{eq:Gamma_gamma}
\Gamma_\gamma
=\frac{g_{a\gamma}^{2}m_{a}^{3}}{64\pi} \simeq 3.3t_{\rm U}^{-1}
\Big[
\frac{g_{a\gamma}}{10^{-7}~{\rm GeV}^{-1}}
\Big]^2
\left(\frac{m_{a}}{\rm eV}\right)^{3},
\end{equation}
has to be compared with the inverse age of the Universe $t_{\rm U}^{-1}\simeq 2.3\times10^{-18}\s^{-1}$, revealing the instability of these particles on the cosmological scale unless their photon coupling is strongly suppressed, i.e.
   \begin{equation}
        g_{a\gamma}\lesssim 5.5\times
        10^{-7}~{\rm GeV}^{-1}\left(\frac{m_{a}}{\rm eV}\right)^{-3/2}\,.
    \end{equation}
In the standard approach, the axion-photon coupling cannot be arbitrarily small because it is mainly set by the Peccei-Quinn (PQ) energy scale, $f_{a}$, considered to be a fundamental parameter of the theory. For example, in the case of QCD axions, $f_{a}$ governs the breaking of the global PQ symmetry generating the axion as associated pseudo-Goldstone boson~\cite{GrillidiCortona:2015jxo}. The PQ scale can be expressed in terms of the axion-photon coupling as
\begin{equation}
    f_{a}\simeq \xi\frac{\alpha}{2\pi g_{a\gamma}}\simeq 
    10^{17}~\GeV\left(\frac{g_{a\gamma}}{10^{-20}~\GeV^{-1}}\right)^{-1}\,,
    \label{eq:PQscale}
\end{equation}
where the model dependent factor $\xi$ is assumed to be $\mathcal{O}(1)$, and $\alpha$ is the fine-structure constant. The requirement for the PQ scale to be below the Planck scale, i.e.~$f_{a}\lesssim M_{\rm P}$ with $M_{\rm P}=1.22\times10^{19}$~GeV, sets a lower limit on $g_{a\gamma}$ and hence on the largest stable axion DM mass scale, $m_{a}\lesssim 30~\GeV$, which are being challenged by strong observational bounds. In order to further suppress $g_{a\gamma}$ and to push for higher axion masses, there are certain attempts to accommodate `super-Planckian axions' (see e.g.~Refs.~\cite{Banks:2003sx,Bachlechner:2014gfa,Fonseca:2019aux}) which however face big challenges to achieve a strong hierarchy $f_a\gg M_{\rm P}$. Therefore, new ideas are required to go significantly beyond the current state-of-the-art to unveil new unexplored avenues in ALP DM research.

The above mentioned challenges of the conventional axion models are deeply rooted in the theoretical interpretation of $f_a$ as a fundamental energy scale associated with PQ symmetry breaking and with the axion being a pseudo-Goldstone boson of that symmetry breaking. We propose a new theoretical framework based upon composite dynamics where $f_a$ appears as an emergent scale which is not associated with any fundamental symmetry breaking, while the ALP itself is no longer considered as a pseudo-Goldstone mode. This framework leads to a new class of composite ALP-like DM candidates naturally reaching a much larger $m_a$, as well as a much smaller $g_{a\gamma}$ and, consequently, much larger emergent (seesaw-like) PQ scale than in conventional ALP models allowing for a strong $f_a\ggg M_{\rm P}$ hierarchy. Below, we demonstrate that such ALP-like states originate in confining Yang-Mills dark sectors feebly coupled to photons.

\emph{\textbf{Dark gluons and glueballs.}---} We postulate the existence of a dark sector with an additional exact $SU(N)$ gauge symmetry, with number of colors $N\ge 3$. The dynamics of the corresponding massless gauge bosons (`dark gluons') is governed by the Lagrangian,
\begin{equation}
\begin{split}
    \mathcal{L}_{\rm SU(N)}&=-\frac{1}{4}G_{\mu\nu}^{a}G^{\mu\nu a}+\frac{\theta }{4}G_{\mu\nu}^{a}\tilde{G}^{\mu\nu a}\,,\\
    G_{\mu\nu}^{a} &= \partial_\mu A_{\nu}^{a} - \partial_\nu A_{\mu}^{a} + g f^{abc} A_{\mu}^{b} A_{\nu}^{c} \,,
    \label{eq:lagrangian}
\end{split}
\end{equation}
with gauge fields $A_{\mu}^{c}$, the coupling $g=g(\mu)$ running with the scale $\mu$, and the structure constants $f^{abc}$, $a,b,c=1,\dots,N^2-1$. Here, $G_{\mu\nu}^{a}$ is the field strength tensor and $\tilde{G}_{\mu\nu}^{a}=\frac{1}{2}\epsilon_{\mu\nu\alpha\beta}G^{\alpha\beta a}$ is its dual. The $\theta$-term implies the Charge-Parity (CP) violation. The Lagrangian~\eqref{eq:lagrangian} features color confinement phenomenon at energies below confinement scale, $\mu\lesssim \Lambda$, such that only composite states neutral under $SU(N)$ appear in the low-energy spectrum~\cite{Politzer:1973fx,Gross:1973ju,Gross:1974cs} (see Refs.~\cite{Yamada:2022imq,Yamada:2022aax} for other groups). 

Confinement allows for the formation of a tower of color-neutral glueballs represented by all possible composite operators that are invariant under $SU(N)$.
In previous studies~\cite{Carenza:2022pjd,Carenza:2023eua}, the authors discussed the cosmological dynamics of scalar glueballs, ${\cal H}\equiv 0^{++}$, in $SU(N)$ theories with $N=3,4,5$. A low-energy effective field theory (EFT) was used to describe this dynamics, where dark gluon effects at high temperatures are characterized by the Polyakov loop~\cite{Sannino:2002wb}. In this work, we incorporate also the pseudoscalar glueball state, ${\cal A}\equiv 0^{-+}$, featuring the properties of standard ALPs, and uncover its important phenomenological implications.

Specifically, the $0^{++}$ and $0^{-+}$ glueball fields are defined in terms of gauge-invariant operators of the lowest dimension as
\begin{equation}
   {\cal H}^{4} \equiv -\frac{\beta(g)}{2g} G^{a}_{\mu\nu}G^{\mu\nu a}\,,\quad
  {\cal A}\, {\cal H}^{3} \equiv G^{a}_{\mu\nu}\tilde{G}^{\mu\nu a}\,,
  \label{GG-to-glue}
\end{equation}
in terms of $\beta$-function $\beta(g)$, whose dynamics is led by the composite EFT Lagrangian~\cite{Rosenzweig:1979ay,Schechter:1980ak}
\begin{equation}
    \mathcal{L}_{\rm eff}=\frac{1}{2}\partial_{\mu}{\cal H}\partial^{\mu}{\cal H}+\frac{1}{2}\partial_{\mu}{\cal A}\partial^{\mu}{\cal A}-V_{\rm eff}({\cal H},{\cal A})\,.
    \label{eq:efflagV}
\end{equation}
Here, the effective glueball potential $V_{\rm eff}$ satisfies the non-conservation of the dilatonic current of the dark gluon sector yielding the trace anomaly relation~\cite{Peskin:1995ev,Collins:1976yq,Rosenzweig:1979ay}
\begin{equation}
    \begin{split}
\partial_{\mu}D^{\mu}&=\frac{\beta(g)}{2g}G_{\mu\nu}^{a}G^{\mu\nu a}=-{\cal H}^4\,,
\label{eq:noncons}
    \end{split}
\end{equation}
being an important consequence of the running of the gauge coupling $g(\mu)$ determined by $\beta(g)$. This relation holds at the non-perturbative level and enables us to constrain the structure of the glueball interactions. The $\theta$-term being a total derivative of the topological current does not receive perturbative quantum corrections \cite{Marino:2015yie,Vicari:2008jw}. Thus, its $\beta$-function vanishes at all orders in perturbation theory yielding no contribution to Eq.~(\ref{eq:noncons}).

From the Lagrangian in Eq.~\eqref{eq:efflagV}, the dilatation current can be calculated for a field $\varphi \equiv \{{\cal H},{\cal A}\}$ that transforms as $\varphi'(x')=\varphi'(\lambda^{-1}x)=\lambda\, \varphi(x)$, leading to 
\begin{equation} 
\partial_{\mu}D^{\mu} = -\Theta^{\mu}_{\mu}=4V_{\rm eff}-\frac{\partial V_{\rm eff}}{\partial {\cal H}}{\cal H}-\frac{\partial V_{\rm eff}}{\partial {\cal A}}{\cal A}\,,
\label{eq:dilaton}
\end{equation}
where $\Theta_{\mu\nu}$ is the canonical energy-momentum tensor. Therefore, comparing Eqs.~\eqref{eq:noncons} and \eqref{eq:dilaton}, the composite EFT potential is restricted to be in the following form
\begin{equation}
\begin{split}
    V_{\rm eff}&\simeq c_{0}{\cal H}^{4}\ln\left(\frac{{\cal H}}{\Lambda}\right)+\frac{\theta}{4}{\cal A}\,{\cal H}^{3}+{\cal H}^{4}f\left(\frac{{\cal A}}{{\cal H}}\right)+\\
&\quad+c_{1}{\cal H}^{4}+c_{2}{\cal H}^{2}{\cal A}^{2}+c_{3}{\cal A}^{4}+c_{4}{\cal H}\,{\cal A}^{3}\,,
\end{split}
    \label{eq:potential}
\end{equation}
in terms of an arbitrary continuous function $f$. Here, the CP-noninvariant terms with odd powers of ${\cal A}$ are permitted due to $\theta\neq0$ in the fundamental theory, see Eq.~\eqref{eq:lagrangian}. The parameters $\theta$, $c_{i}$, $i=0,\dots,4$ and the function $f$ can, in principle, be fully determined by lattice simulations.

Furthermore, the potential in Eq.~\eqref{eq:potential} can be expanded in power series over glueball excitations $\eta\equiv {\cal H}-\eta_{0}$ and $a\equiv {\cal A}-a_{0}$ around its minimum located at the vacuum expectation values (VEVs) of the glueball fields $\{\eta_{0}, a_{0}\}\sim\mathcal{O}(1)\Lambda$. Then, turning to the mass basis of these excitations by means of rotation by an angle $\delta$, the resulting EFT potential can be represented as follows,
\begin{equation}
  V_{\rm eff} \simeq \frac{m_{1}^{2}}{2} \phi_1^{2} + \frac{m_{2}^{2}}{2}\phi_2^{2}+\sum_{i=0}^{\infty}\sum_{j=0}^{\infty}\frac{\lambda_{ij}}{\Lambda^{i+j-4}}\phi_1^{i} \phi_2^{j}\Bigg|_{i+j\ge3}\,,
  \label{eq:exp}
\end{equation}
where $\phi_1\equiv c_\delta\eta - s_\delta a$ and $\phi_2\equiv s_\delta\eta + c_\delta a$ are the physical glueball fields, with masses $m_{1}$ and $m_{2}$, respectively, with $c_\delta\equiv \cos\delta$ and $s_\delta\equiv \sin\delta$, while $\lambda_{ij}$ are dimensionless coefficients. The higher-dimensional operators for $i+j>4$ in the above potential are suppressed by increasing powers of $\Lambda$, hence, can be ignored at low energies.

\emph{\textbf{Glueball dark matter.}---} Dark glueballs are often considered in the literature as viable DM candidates~(see e.g.~Refs.~\cite{Carenza:2022pjd,Carenza:2023eua,Carlson:1992fn,Faraggi:2000pv,Feng:2011ik,Boddy:2014yra,Soni:2016gzf,Kribs:2016cew,Acharya:2017szw,Dienes:2016vei,Soni:2016yes, Soni:2017nlm,Draper:2018tmh,Halverson:2018olu, Forestell:2017wov, Forestell:2016qhc,Yamada:2023thl,McKeen:2024trt,Biondini:2024cpf} and references therein). Starting with the two-glueball EFT formulated above, let us focus on the mass hierarchy satisfying~$m_1/2 < m_2 < 2m_1$ such that both glueball species $\phi_{1,2}$ would be stable and contribute to the DM abundance after their freeze out. By naive dimensional counting, in the strong coupling regime the glueball self-interactions happen with rates of order $\Lambda$ being extremely efficient on cosmological scales close to the dark confinement phase transition epoch at the critical temperature $T_c\sim \Lambda$, i.e.~$\Lambda\gg H(T_c)$. Both glueball species freeze-out with respect to particle number changing processes $m\to n$, with the $3\to2$ scattering being the last one to become inefficient. This occurs soon after the confinement phase transition when the glueballs are non-relativistic at temperature $T_f\lesssim T_c$. Since then the resulting total DM density is impacted by cosmological dilution only. The relative abundance of $\phi_{1,2}$ may be roughly estimated in terms of their mass splitting $\Delta m = m_1 - m_2$ as $n_1/n_2 \sim (m_1/m_2)^{-3/2}\exp(-\Delta m/T_f)> 10^{-3}$ that holds at freeze-out and is preserved until today. The exact ratio $n_1/n_2$ is not relevant for generic properties of the glueball DM as will be discussed below.

As demonstrated in Refs.~\cite{Carenza:2022pjd,Carenza:2023eua}, the first-order phase transition in the glueball system washes out any significant dependence of the relic density on the initial conditions. In addition, the energy of the dark sector in the comoving volume is conserved since the rates of glueball electromagnetic interactions $\Gamma_\gamma$ are well below the Hubble expansion rate $H$ at $T<T_c$, and hence interactions between the dark and visible sectors can be safely ignored. These two conditions imply that the total glueball DM relic density~\cite{Carenza:2022pjd,Carenza:2023eua},
\begin{equation}
        \Omega_{\rm DM}h^{2}\simeq 0.12\,\zeta_{T}^{-3}\frac{\Lambda}{\Lambda_{0}}\,,\quad 100~\eV\lesssim\Lambda_0\lesssim 400~\eV \,,
    \label{eq:lambda0}
\end{equation}
with visible-to-dark sector temperature ratio $\zeta_{T}\equiv T_\gamma/T$, holds approximately in the considered two-glueball DM framework scaling linearly with the confinement scale $\Lambda$ which is constrained to be in a very wide range~\cite{Carenza:2023eua}
\begin{equation}
20~\MeV \lesssim \Lambda \lesssim 10^{10}~\GeV\,.
\label{Lambda-const}
\end{equation}
Here, the upper limit is due to the requirement that the glueballs constitute the totality of DM while the phase transition does not happen before inflation, thus, avoiding exponential dilution of DM. The lower limit applies only in the case of glueballs constituting the ballpark of DM for $\zeta_{T}^{-1}\lesssim0.01$ and is dictated by avoiding too strong DM self-interactions that would affect the innermost regions of galaxy clusters, ultimately in contrast with observations of their surrounding hot plasma~\cite{Eckert:2022qia}. Indeed, the self-interaction GALP cross section $\sigma_{\rm SI}$ is determined by the GALP mass and satisfies $\sigma_{\rm SI}<\sigma_{\rm geom}\sim 1/m_{\rm GALP}^{2}$~\cite{Smirnov:2019ngs}. We also emphasize that for $\Lambda<20~\MeV$, only a small fraction of DM can be in the form glueballs. Note, the linear scaling of~(\ref{eq:lambda0}) highlights the long-standing dark glueball overabundance problem at large $\Lambda$~\cite{Halverson:2016nfq} that can only be alleviated by suppressing the dark-to-visible sector temperature ratio $\zeta_{T}^{-1}$. In a limiting case of thermal equilibrium between visible and dark sectors, i.e.~$\zeta_{T}^{-1}\to 1$, the confinement scale $\Lambda\to\Lambda_0$ is constrained to low values within a narrow range in Eq.~(\ref{eq:lambda0}). Note that $\zeta_{T}^{-1}<0.37$ is required to avoid cosmological bounds on dark radiation~\cite{Carenza:2022pjd,Carenza:2023eua}.

\emph{\textbf{Glueball-photon interactions.}---} In order to introduce feeble interactions between the visible and dark QCD-like sectors, we adopt the minimal QCD-like framework with a single heavy Dirac fermion, $\Psi$, acting as a portal coupled to both dark gluons and photons. At variance with standard QCD, such a fermion is assumed to have a very large mass $M_\Psi \gg \Lambda$ that implies its negligible impact on the confinement dynamics and, hence, on the composite glueball EFT discussed above. Furthermore, the effective interactions between dark gluons and photons are given by higher dimensional operators generated at one-loop level by integrating out $\Psi$ in the loop. Thus, the strength of electromagnetic interactions of glueballs is suppressed and is controlled by a fermion mass $M_\Psi$ and the gauge coupling at that scale, $g(M_\Psi)$.

Consider the fermion $\Psi$ with an electric charge $q_{\Psi}$ (in units of $e$). Then, the dark SU(N) gauge coupling at the UV $\mu=M_\Psi$ scale can be written as $g(\mu)\equiv \tau\,e(\mu)q_{\Psi}$. Assuming no additional new physics below the UV cutoff scale, we find $\tau\sim 1$ following the one-loop RG evolution with $N=3$ and $q_{\Psi}=1$, under the requirement that the IR Landau pole is located close to the confinement scale $\Lambda\ll M_\Psi$. Consequently, relevant dimension-8 operators read~\cite{Faraggi:2000pv,
Juknevich:2009ji}
\begin{equation}
    \mathcal{L}_{\rm eff} \supset \frac{\tau^2\alpha^2}{M_\Psi^4}\,
    \Big[c_\gamma\,G_{\mu\nu}^{a}G^{\mu\nu a} F_{\alpha\beta}F^{\alpha\beta} + \tilde{c}_\gamma\, G_{\mu\nu}^{a}\tilde{G}^{\mu\nu a} F_{\alpha\beta}\tilde{F}^{\alpha\beta} \Big]\,,
    \label{eq:naive}
\end{equation}
where $\alpha\equiv\alpha(M_\Psi)=e(M_{\Psi})^2/4\pi$ is the fine structure constant, $F^{\alpha\beta}$ and $\tilde{F}^{\alpha\beta}$ are the photon field strength tensor and its dual, respectively, while dimensionless $c_\gamma$ and $\tilde{c}_\gamma$ can be evaluated perturbatively at the scale $M_\Psi$. The EFT description in Eq.~(\ref{eq:naive}) is valid as long as $\Lambda \ll M_\Psi$. As soon as the temperature drops below $T_c\sim \Lambda$, one turns to composite EFT. Using Eq.~(\ref{GG-to-glue}), then expanding the glueball fields about their VEVs, and keeping linear terms in $a$ and $\eta$ only, we get
\begin{equation}
G_{\mu\nu}^{a}G^{\mu\nu a} = -\frac{8g\eta_0^3\,\eta}{\beta(g)}\,, \quad
G_{\mu\nu}^{a}\tilde{G}^{\mu\nu a} = \eta_0^3a + 3a_0\eta_0^2\eta\,,
\end{equation}
where $g/\beta(g)$ should be computed in the non-perturbative regime of the dark Yang-Mills theory at $\mu\sim \Lambda$ corresponding to a large $g(\mu)\gg 1$. According to the recent lattice results \cite{Hasenfratz:2023bok}, in this regime $g/\beta(g)\sim 1$. Turning to the mass basis $\{\eta,a\} \to \{\phi_1,\phi_2\}$, the glueball-photon interactions are governed by the Lagrangian 
\begin{equation}
    \mathcal{L}_{\rm eff} \supset \frac{1}{4}\sum_{i=1,2}\Big[g_{\phi_i\gamma}\, \phi_i F_{\mu\nu}F^{\mu\nu} + \tilde{g}_{\phi_i\gamma}\, \phi_i F_{\mu\nu}\tilde{F}^{\mu\nu} \Big]\,.
    \label{eq:lagh}
\end{equation}

\emph{\textbf{GALP mass-coupling relation}.---} A strong CP-violation in the dark sector implies that the glueball VEVs satisfy $\eta_0\sim a_0\sim \Lambda$ while the $\eta$-$a$ mixing angle $\delta$ is large. As a result, from the viewpoint of particle astrophysics the physical glueballs $\phi_{1,2}$ interact with photons with similar rates while having similar masses effectively acting as axion DM. Hence, we refer to these states as Glueball Axion-Like Particles (GALPs) whose effective couplings satisfy $g_{\phi_i\gamma} \sim \tilde{g}_{\phi_i\gamma} \equiv g_{{\rm GALP}\gamma}$, where
\begin{eqnarray}
     &&g_{{\rm GALP}\gamma} = \kappa\,\alpha^{2} \Lambda^{-1}\Big[ \frac{\Lambda}{M_\Psi}\Big]^{4} \label{eq:gag} \\
     &&\quad =\,2.45\times10^{-7}\GeV^{-1}\kappa\Big[\frac{m_{\rm GALP}}{\GeV}\Big]^{3} \Big[\frac{M_\Psi}{\GeV}\Big]^{-4} \,.\nonumber
\end{eqnarray}
Here, as a proof-of-concept, we introduce a universal GALP mass scale $m_1\sim m_2\equiv m_{\rm GALP}=6\Lambda$~\cite{Curtin:2022tou}, and a universal dimensionless coefficient $\kappa$ expected to be of order 0.1 -- 10. The relation (\ref{eq:gag}) is a defining property for the whole class of novel GALP DM scenarios featuring a specific dependence on two distinct scales: $m_{\rm GALP}\ll M_\Psi$ and $M_\Psi$. This can be compared to an analogous relation for QCD axions, e.g.~\mbox{$g_{a\gamma}=2\times10^{-10}\GeV^{-1}(m_{a}/\eV)$} for the case of Kim-Shifman-Vainshtein-Zakharov axions~\cite{GrillidiCortona:2015jxo,Gorghetto:2018ocs}.

Given a remarkable similarity between GALPs and standard axions, it is instructive to determine an effective PQ-like scale by plugging Eq.~\eqref{eq:gag} into~\eqref{eq:PQscale}
\begin{equation}
    f_{a}\simeq4.1\times10^{3}~{\rm GeV}\,\kappa^{-1}\Big[\frac{\Lambda}{\GeV}\Big]^{-3}\Big[\frac{M_\Psi}{\GeV}\Big]^{4}\,.
\end{equation}
It is evident that this emergent scale can be well in the super-Planckian domain that is considered impossible to reach in any other model adopting it as a fundamental scale, e.g.~in string theory $f_{a}\lesssim 10\, M_{\rm P}$~\cite{Banks:2003sx,Bachlechner:2014gfa}.

In the minimal EFT approach developed above, the GALP  coupling to photons, $g_{{\rm GALP}\gamma}$, is linked to the GALP mass, $m_{\rm GALP}$. Let us investigate how this relation correlates with the DM relic abundance. Assume that for sufficiently high temperatures $T\gg M_{\Psi}$ the SM bath and the dark-sector $\Psi$ fermion are in thermal equilibrium. As the Universe cools down to $T\sim M_{\Psi}$, $\Psi$ and $\bar{\Psi}$ start to annihilate into the SM $X_{\rm SM}$ and dark gluon $\tilde{g}\tilde{g}$ final states, with rates $\Gamma_{\Psi\bar{\Psi}\to X_{\rm SM}}$ and $\Gamma_{\Psi\bar{\Psi}\to \tilde{g}\tilde{g}}$, respectively. For these processes to efficiently reheat the SM bath relative to the dark sector, hence to suppress the dark-to-visible temperature ratio, $\zeta_T^{-1}$, we introduce a small parameter
\begin{equation}
    \epsilon \equiv \frac{\Gamma_{\Psi\bar{\Psi}\to \tilde{g}\tilde{g}}}{\Gamma_{\Psi\bar{\Psi}\to X_{\rm SM}}} \ll 1 \,,
\end{equation}
such that $\Psi$-fermions annihilate mostly into the SM particles. This parameter $\epsilon$ is a key observable characteristic of the portal interactions between the dark and visible sectors and depends on the details of the UV model.

At the temperature of the dark sector, $T\equiv T_{\tilde{g}} \simeq M_\Psi$, when $\Psi\bar{\Psi}$-annihilation completes, the resulting densities of photons $\rho_{\gamma}$ and dark gluons $\rho_{\tilde{g}}$ are related as
\begin{equation}
    \frac{\rho_{\tilde{g}}}{\rho_{\gamma}} \sim \epsilon \sim (N^{2}-1)\Big(\frac{T}{T_{\gamma}}\Big)^4 \,,
\end{equation}
where the prefactor indicates the number of dark gluon degrees of freedom. Then, as the Universe cools down to the modern temperature $T^{(0)}_{\gamma}$, the entropy production events in the visible sector also affect the photon temperature increasing it by an additional factor of $\big[g_{*,s}(T^{(0)}_{\gamma})/g_{*,s}(M_\Psi)\big]^{-1/3}$. Here, $g_{*,s}(T^{(0)}_{\gamma})=3.909$ is the number of entropic degrees of freedom of the relativistic (photon) bath today, while $g_{*,s}(M_\Psi)$ refers to the same quantity at the $\Psi\bar{\Psi}$-annihilation epoch (equal to 106.75 for pure SM). As a result, the dark-to-visible sector temperature ratio in the modern epoch reads,
\begin{equation}
    \zeta_{T}^{-1}=\left(\frac{\epsilon}{N^{2}-1}\right)^{1/4}\Big[\frac{g_{*,s}(T^{(0)}_{\gamma})}{g_{*,s}(M_\Psi)}\Big]^{1/3} \ll 1\,.
\end{equation}
The condition for GALPs constituting the totality of DM, i.e.~$\zeta_{T}^{-3}m_{\rm GALP}/(6\Lambda_0)\simeq 1$, effectively connects the dark-to-visible portal information encoded in $\epsilon$ with the GALP mass. In numerical analysis we explore the parameter space $\{g_{{\rm GALP}\gamma},m_{\rm GALP}\}$ particularly relevant for DM searches setting bounds on it for the first time by taking $\kappa\sim 1$ without any loss of generality. We note that it is straightforward to rescale the allowed values of $g_{{\rm GALP}\gamma}$ by a suitable value of $\kappa$ once it becomes available from lattice simulations. For a direct connection of the GALP-to-photon decay rate with the non-perturbative matrix elements that would be probed by lattice calculations, see Ref.~\cite{Juknevich:2009ji}.
\begin{figure*}[t!]
    \vspace{0.cm}
    \includegraphics[width=0.99\linewidth]{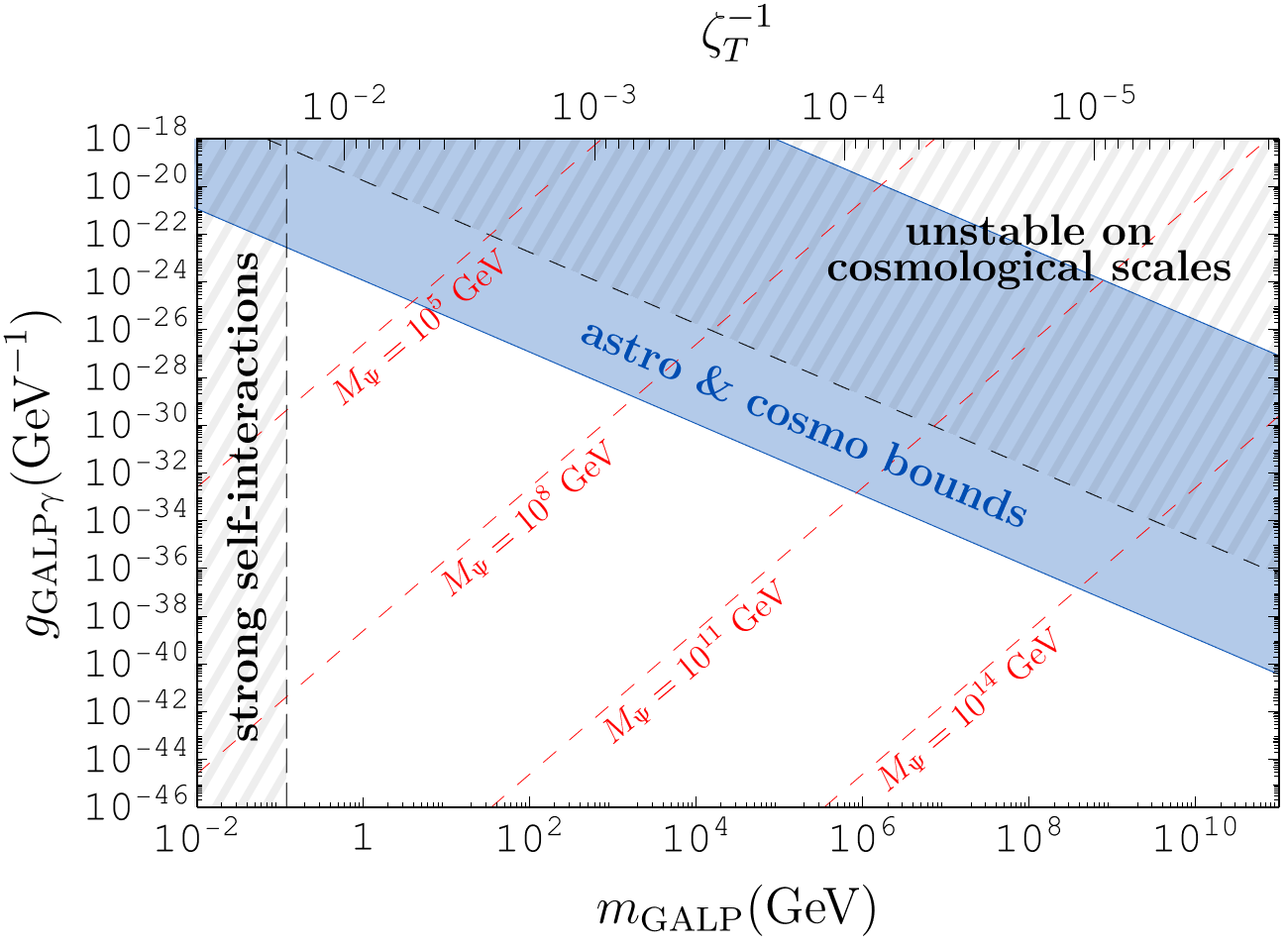}
    \caption{The GALP-photon coupling $g_{{\rm GALP}\gamma}$ versus mass $m_{\rm GALP}=6\Lambda$. The upper hatched  domain is excluded by the GALP DM stability constraint, requiring GALP stability on cosmological scales $\Gamma_{\gamma}<t_{\rm U}^{-1}$;  the left hatched domain is excluded by bounds on self-interacting DM~\cite{Carenza:2022pjd},  while the blue region is ruled out by astrophysical DM searches as well as by the cosmological bound specified in the text. The red dashed lines represent the GALP coupling-mass relation in Eq.~\eqref{eq:gag} for four distinct values of $M_\Psi$. The upper axis shows the value of $\zeta_{T}^{-1}$ needed to make GALPs the totality of DM for a given mass, considering $\Lambda_{0}=100$~eV.}
    \label{fig:models}
\end{figure*}

\emph{\textbf{Discussion and conclusions.}---} In Fig.~\ref{fig:models} we show the parameter space of GALPs constituting the totality of DM, $\{g_{{\rm GALP}\gamma},m_{\rm GALP}\}$ being represented by Eq.~\eqref{eq:gag}. The upper hatched domain is excluded due to the cosmological bound of the GALP stability while the left hatched domain for light $m_{\rm GALP}\gtrsim 120$~MeV is due to DM self-interaction constraint~\cite{Smirnov:2019ngs}. GALPs decaying into photons during the Big Bang Nucleosynthesis would affect the primordial nuclei abundances. We extended the bound of~\cite{Depta:2020wmr} towards the higher mass limiting it to be below the reheating temperature which gives rise to the upper bound of the excluded blue domain. Its lower bound represents the constraint from the ultra-massive decaying DM searches~\cite{Blanco:2018esa,Munbodh:2024ast} reaching the mass of up to $10^{11}$~GeV and the decay rates of down to $\sim\mathcal{O}(10^{-26}{\rm s}^{-1})$, or even lower. The slices of the parameter space for each fixed value of $M_\Psi$ are represented by red dashed lines. Note that above the EFT boundary $m_{\rm GALP} \simeq M_\Psi$ the considered EFT approach breaks down. Thus, the phenomenologically viable (white) domain is in agreement with the existing constraints.

Due to GALP DM being heavy with $\Psi$-loop suppressed interactions to the visible sector, the unitarity constraint~\cite{Griest:1989wd}, required by DM freeze-out of thermal equilibrium with the SM bath, can be naturally avoided. GALPs were not produced by freeze out, but via confinement in the dark gauge sector, while equilibrium with visible one was reached only asymptotically at $T\gtrsim M_\Psi\gg \Lambda$, i.e.~in deconfined phase. The GALPs were formed at $T\equiv \zeta_T^{-1}T_\gamma \sim \Lambda$ with $\zeta_T^{-1}\ll 1$, such that they were not in equilibrium with the SM bath since their formation until the present epoch. This can be seen by comparing the GALP-photon interaction rate
$\Gamma_{\gamma}\sim g_{{\rm GALP}\gamma}^{2}T_{\gamma}^{3}\zeta^{-3}_{T}$ with $H\sim T_{\gamma}^{2}/M_{\rm P}$, leading to the equilibrium condition $g_{{\rm GALP}\gamma}\gtrsim (T_{\gamma}\zeta^{-3}_{T} M_{\rm P})^{-1/2}$. The latter suggests that GALPs with $g_{{\rm GALP}\gamma}\lesssim 10^{-18} \GeV^{-1}$ have not been in thermal equilibrium since their production epoch, in agreement with~\cite{Cadamuro:2011fd}. Fig.~\ref{fig:models} shows large domains of the yet unexplored parameter space for ultra-heavy DM GALPs which are expected to be probed by indirect DM searches~\cite{Cirelli:2010xx,Cirelli:2024ssz}.

In conclusions, our work provides an innovative framework suggesting a whole new class of composite DM models featuring ALP-type interactions with photons characterized by the novel electromagnetic coupling-mass relation~(\ref{eq:gag}). In the suggested framework, ALPs emerge as glueball states (or GALPs) of a confining dark Yang–Mills sector. We build the first minimal realisation of GALPs with yet unexplored vast domains of the physical parameter space naturally featuring large masses $m_{\rm GALP} > 120~\mathrm{MeV}$ and highly suppressed photon couplings $g_{{\rm GALP}\gamma} < 10^{-23}~\mathrm{GeV}^{-1}$ that arise without any fine-tuning due to an emergent PQ-like scale. The minimality of the GALP DM framework is two-fold: on the UV theory side, it contains just an extra confining dark gauge symmetry and a heavy fermion $\Psi$, while on the low-energy EFT side, it exhibits only three phenomenologically relevant parameters entering~(\ref{eq:gag}): $m_{\rm GALP}\simeq 6\Lambda$, $M_\Psi\gg \Lambda$, $\kappa\sim 0.1-10$. The unique feature of GALPs stems from the dark gluon condensation phenomenon associated with confinement leading to distinct two-scale (seesaw-like) dependence $g_{{\rm GALP}\gamma}\propto m_{\rm GALP}^{3}/M_\Psi^4$ unattainable for standard ALPs characterized by linear mass scaling, $g_{a\gamma}\propto m_a$. Our pioneering studies suggest that the heavy ALP DM research could become one of the important new avenues in astrophysics in near future. Finally, our work comes as an important motivation for the lattice community to analyze the glueball self-interactions and thermal evolution reaching the next milestone in exploration of composite dynamics.

\emph{Acknowledgments.---} We warmly thank Pedro de la Torre Luque and  John March-Russell for fruitful discussions. The work of P.C. is supported by the European Research Council under Grant No. 742104 and by the Swedish Research Council (VR) under grants 2018-03641, 2019-02337 and 2022-04283. This article is based upon work from COST Action COSMIC WISPers CA21106, supported by COST (European Cooperation in Science and Technology). Z.-W.W. is supported in part by the National Natural Science Foundation of China (Grant No.~12475105).

\bibliographystyle{bibi}
\bibliography{biblioPRL.bib}

\end{document}